\documentclass[a4paper]{jpconf}
\usepackage{graphicx}

\begin{document}

\title{Transport coefficients in second-order non-conformal viscous hydrodynamics}

\author{Radoslaw Ryblewski}

\address{The H. Niewodnicza\'nski Institute of Nuclear Physics, Polish Academy of Sciences, PL-31342 Krak\'ow, Poland}

\ead{Radoslaw.Ryblewski@ifj.edu.pl}

\begin{abstract}
Based on the exact solution of Boltzmann kinetic equation in the relaxation-time approximation, the precision of the two most recent formulations of relativistic second-order non-conformal viscous hydrodynamics (14-moment approximation and causal Chapman-Enskog method), standard Israel-Stewart theory, and anisotropic hydrodynamics framework, in the simple case of one-dimensional Bjorken expansion, is tested. It is demonstrated that the failure of Israel-Stewart theory in reproducing exact solutions of the Boltzmann kinetic equation occurs due to neglecting and/or choosing wrong forms of some of the second-order transport coefficients. In particular, the importance of shear--bulk couplings in the evolution equations for dissipative quantities is shown. One finds that, in the case of the bulk viscous pressure correction, such coupling terms are as important as the corresponding first-order Navier-Stokes term and must be included in order to obtain, at least qualitative, overall agreement with the kinetic theory.
\end{abstract}

\section{Introduction}
The theory of relativistic hydrodynamics turned out to be very successful in describing soft hadronic observables produced in ultra-relativistic heavy-ion collisions at RHIC and the LHC \cite{Bozek:2009ty, Bozek:2012qs}; for a review see also \cite{Florkowski:2010zz}. Due to existence of the lower bound of the shear viscosity to entropy density ratio resulting from quantum mechanical considerations, dissipative effects must be included in order to realistically model the quark-gluon plasma \cite{Kovtun:2004de}. However, despite the successes of the first and most widely used, Israel-Stewart, formulation of causal relativistic second-order viscous hydrodynamics \cite{Israel:1979wp}, the latter is known to suffer from certain approximations and assumptions. As a result, there are recently multiple developments focusing on constructing more complete and self-consistent methods for deriving the fluid-dynamical equations of motion. These methods include complete second-order treatments \cite{Denicol:2012cn,Denicol:2014vaa}, third-order treatments \cite{El:2009vj,Jaiswal:2013vta} and the anisotropic hydrodynamics framework \cite{Martinez:2010sc,Florkowski:2010cf,Martinez:2010sd,Ryblewski:2010bs,
Ryblewski:2011aq,Florkowski:2011jg,Strickland:2012bc,Florkowski:2012lba,
Martinez:2012tu,Ryblewski:2012rr,Florkowski:2012as,Florkowski:2012pf,
Rybczynski:2012ee,Florkowski:2012mz,Florkowski:2013uqa,Tinti:2013vba,
Bazow:2013ifa,Strickland:2013uga,Strickland:2014eua,Florkowski:2014bba,
Nopoush:2014pfa,Heinz:2014zha,Strickland:2014pga,Nopoush:2014qba,
Ryblewski:2013jsa}. The forms of the transport coefficients that are entering all of the dissipative hydrodynamics formulations have to be determined by matching the latter with the underlying microscopic kinetic theory. As a result, the minimal requirement for a reasonable non-conformal hydrodynamic theory is that it should be able to describe the evolution of a viscous medium at least in this regime. 

In principle, all of the hydrodynamic studies based on the kinetic Boltzmann equation use the so-called relaxation-time approximation for the collisional kernel. Recently it was shown that in this case the exact solution of the Boltzmann kinetic equation can be constructed providing one restricts oneself to the case of Bjorken flow \cite{Florkowski:2013lza,Florkowski:2013lya,Florkowski:2014sfa,
Florkowski:2014txa,Florkowski:2014cqa}; for generalization to Gubser flow see \cite{Denicol:2014tha,Denicol:2014xca}. In the following we will show that this tool can be extremely useful for testing various dissipative hydrodynamics formulations available in literature against the underlying kinetic theory. We will show that among the cases considered herein, the Chapman-Enskog method leads to the best overall description of the medium in the general, non-conformal, case.

\section{Relativistic dissipative hydrodynamics from kinetic theory}

The evolution of the system within relativistic viscous hydrodynamics is governed by the energy and momentum continuity equations, $\partial_\mu T^{\mu \nu} = 0$, with the energy-momentum tensor given by
\begin{equation}\label{NTD}
T^{\mu\nu} = \epsilon u^\mu u^\nu - (P+\Pi)\Delta^{\mu\nu} + \pi^{\mu\nu},
\end{equation}
where $\epsilon$, $P$, $\Pi$ and $\pi^{\mu\nu}$ are the energy density, pressure, bulk pressure correction and shear-stress tensor, respectively, and $\Delta ^{\mu \nu}\equiv g^{\mu \nu }-u^{\mu }u^{\nu }$ is the projection operator orthogonal to the fluid four-velocity $u^{\mu }$.
One possible way to derive the remaining evolution equations for the dissipative quantities is to start with the underlying Boltzmann kinetic equation $p \cdot \partial f = {\cal C}[f]$. In the case where the collisional kernel is treated in the relaxation-time approximation, ${\cal C}[f] = - (p \cdot u) (f - f_{
eq}) / \tau_{eq}$, energy-momentum conservation equation (\ref{NTD}) must be supplemented by the following relaxation-type equations \cite{Denicol:2014mca,Jaiswal:2014isa}, 
\begin{eqnarray}
\tau_{\Pi} \dot{\Pi} + \Pi &=& 
-\zeta\theta 
-\delta_{\Pi\Pi}\Pi\theta
+\lambda_{\Pi\pi}\pi^{\mu\nu}\sigma_{\mu \nu }, \label{BULK}\\
\tau_{\pi} \dot{\pi}^{\langle\mu\nu\rangle}  + \pi^{\mu\nu} &=&
2\eta\sigma^{\mu\nu}
+2\tau_{\pi}\pi_{\gamma}^{\langle\mu}\omega^{\nu\rangle\gamma}
-\tau_{\pi\pi}\pi_{\gamma}^{\langle\mu}\sigma^{\nu\rangle\gamma}  -\delta_{\pi\pi}\pi^{\mu\nu}\theta 
+\lambda_{\pi\Pi}\Pi\sigma^{\mu\nu}, \label{SHEAR}
\end{eqnarray}
where $\tau_\pi$ and $\tau_\Pi$ are the shear and bulk relaxation time,  $\eta$ and $\zeta$ are the shear and bulk viscosities,  $\omega ^{\mu \nu }\equiv (\nabla ^{\mu
}u^{\nu }-\nabla ^{\nu }u^{\mu })/2$ is vorticity tensor,  \mbox{$\sigma^{\mu%
\nu}\equiv \nabla^{\langle\mu} u^{\nu\rangle}$} is the velocity stress tensor and  $%
\theta \equiv \nabla _{\mu }u^{\mu }$ is the expansion scalar. We use the notation where $\nabla ^{\mu }\equiv \Delta ^{\mu }_{\nu }\partial ^{\nu
} $ is the projected spatial gradient and $A^{\langle \mu
\nu \rangle }\equiv \Delta _{\alpha \beta }^{\mu \nu }A^{\alpha \beta }$,
with $\Delta _{\alpha \beta }^{\mu \nu }\equiv (\Delta _{\alpha }^{\mu
}\Delta _{\beta }^{\nu }+\Delta _{\beta }^{\mu }\Delta _{\alpha }^{\nu
}-2/3\Delta ^{\mu \nu }\Delta _{\alpha \beta })/2$. In Eqs.~(\ref{BULK}) and (\ref{SHEAR}) we also use shorthand notation for the proper-time derivative $\dot{(\,\,)}\equiv d/d\tau $. The quantities $\delta_{\Pi\Pi}$, $\lambda_{\Pi\pi}$, $\tau_{\pi\pi}$, $\delta_{\pi\pi}$ and $\lambda_{\pi\Pi}$ are second-order transport coefficients whose form depends on the method used to derive the evolution equations (\ref{BULK}) and (\ref{SHEAR}). Hereafter, we will consider three methods which are available in literature: Israel-Stewart theory \cite{Israel:1979wp}, 14-moment approximation \cite{Denicol:2014vaa} and Chapman-Enskog method \cite{Jaiswal:2013vta,Jaiswal:2013npa}. Each of these formulations predict the same general form of equations (\ref{BULK}) and (\ref{SHEAR}), however, they lead to different forms of the second-order kinetic coefficients. In the following we will show which method provides the best agreement with the underlying kinetic theory equation in the simple case of purely-longitudinal boost-invariant (Bjorken) expansion. In this case, the results of the viscous hydrodynamics equations may be confronted with the exact solution of the Boltzmann kinetic equation \cite{Florkowski:2013lza,Florkowski:2013lya,Florkowski:2014sfa}. Moreover, in this case, the six independent equations in (\ref{BULK}) and (\ref{SHEAR}) for dissipative fluxes reduce to two differential equations for bulk pressure correction $\Pi$ and rapidity--rapidity coefficient of the shear-stress tensor $\pi^{\mu\nu}$.
\section{Comparison with the exact solution of Boltzmann kinetic equation}

\begin{figure}[t]
\begin{minipage}{18pc}
\includegraphics[width=20pc]{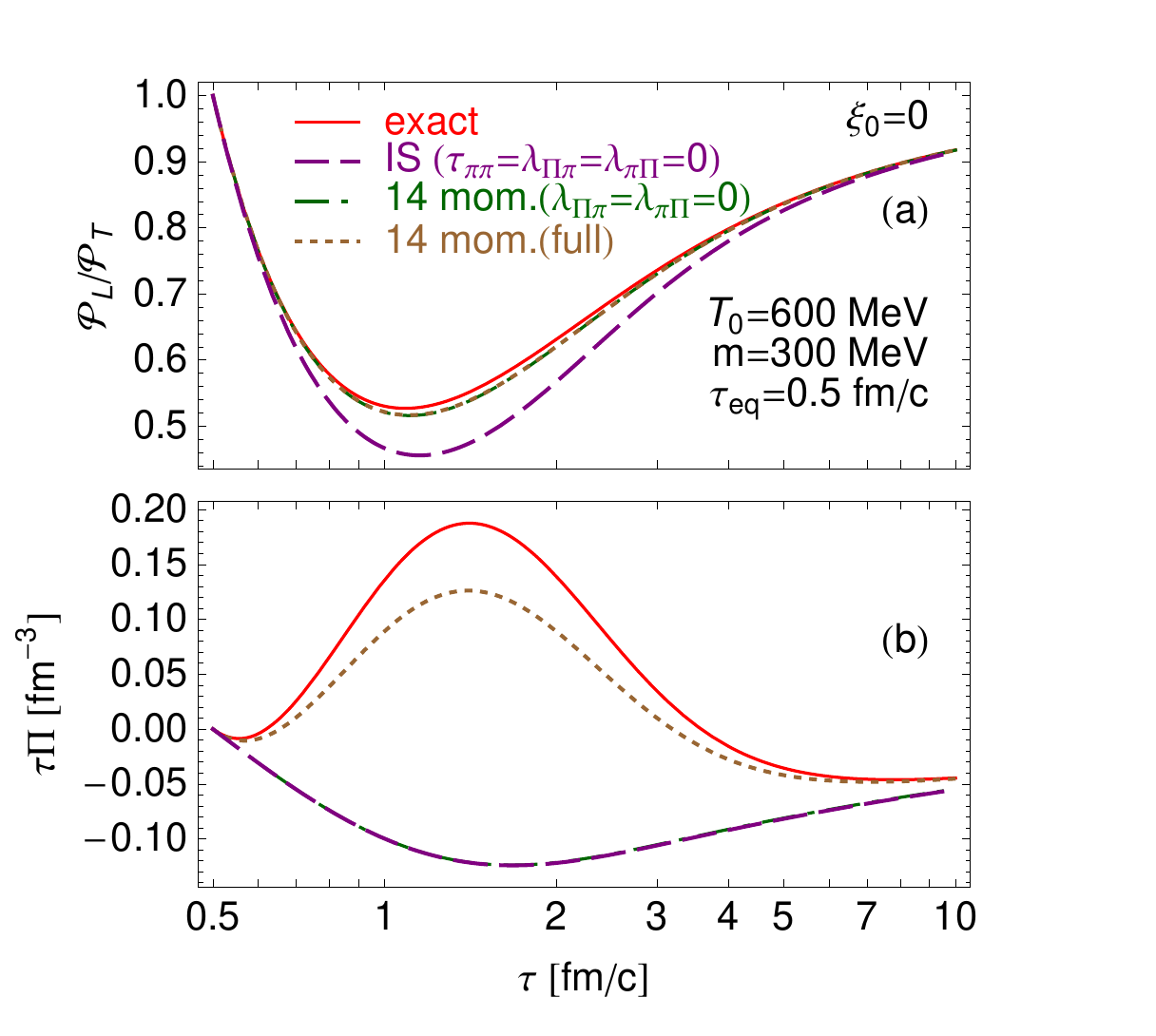}
\caption{\label{fig1} (Color online) Time evolution of (a) pressure anisotropy  and (b) bulk pressure correction obtained using exact solution of the Boltzmann equation (red solid lines), Israel-Stewart viscous hydrodynamics (purple dashed lines), 14-moment approximation without (green dashed-dotted lines) and with (brown dotted lines) shear--bulk couplings.}
\end{minipage}\hspace{1pc}
\begin{minipage}{18pc}
\includegraphics[width=20pc]{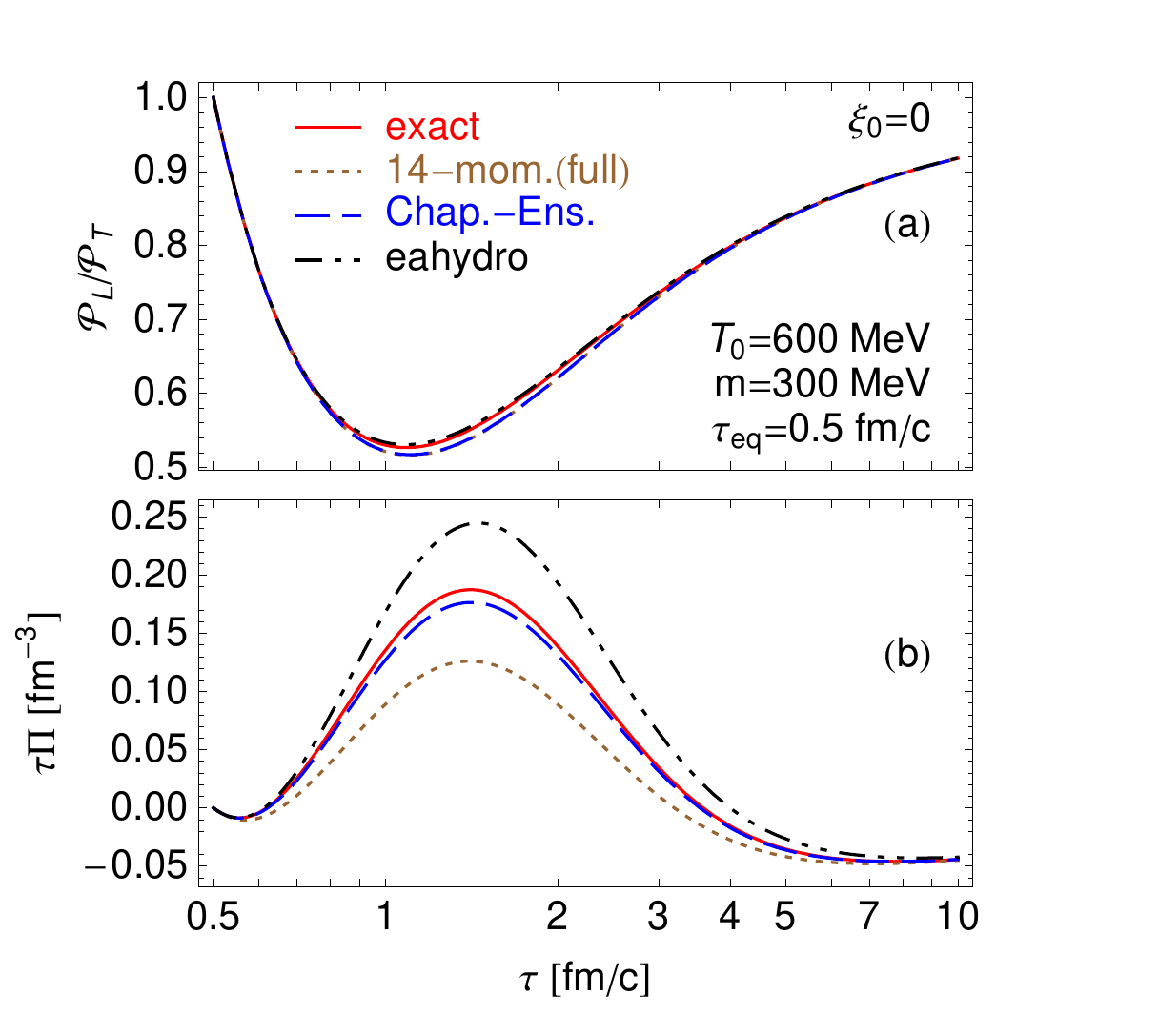}
\caption{\label{fig2} (Color online) Time evolution of (a) pressure anisotropy  and (b) bulk pressure correction  obtained using exact solution of the Boltzmann equation (red solid lines), 14-moment approximation (brown dotted lines), Chapman-Enskog method (blue dashed lines) and anisotropic hydrodynamics (black dashed-dotted-dotted lines).}
\end{minipage} 
\end{figure}

In Fig.~\ref{fig1} we present (a) the pressure anisotropy ${\cal P}_L/{\cal P}_T\equiv (P+\Pi-\pi)/(P+\Pi+\pi/2)$ (directly related to shear stress tensor since $\pi\equiv-\tau^2\pi^{\eta\eta}$) and (b) bulk pressure correction scaled by proper time resulting from two second-order non-conformal viscous hydrodynamics formulations: Israel-Stewart \cite{Israel:1979wp} (purple dashed lines) and 14-moment approximation \cite{Denicol:2014vaa}, with (brown dotted lines) and without (green dashed-dotted lines) shear--bulk couplings. One can observe that the results obtained within traditional Israel-Stewart theory provide the poorest description of  the exact solution of the underlying Boltzmann kinetic equation \cite{Florkowski:2013lza,Florkowski:2013lya,Florkowski:2014sfa} (red solid lines), for both shear and bulk viscous corrections. One observes also that, in order to obtain at least a qualitative description of the bulk pressure correction, one is forced to include the shear--bulk couplings ($\lambda_{\Pi\pi}$ and $\lambda_{\pi\Pi}$) in the two relaxation equations (\ref{BULK}) and (\ref{SHEAR}) for the dissipative quantities.

In Fig.~\ref{fig2}, in addition to the previous results from 14-moment approximation (brown dotted lines) and the exact solution (red solid lines), we also show the results of the alternative formulation of second-order causal viscous hydrodynamics using Chapman-Enskog iterative method \cite{Jaiswal:2013npa,Jaiswal:2013vta} (blue dashed lines) and, for completeness, a novel formulation of anisotropic hydrodynamics framework \cite{Nopoush:2014pfa} (black dashed-dotted-dotted lines). We can immediately see that, while the anisotropic hydrodynamics describes the bulk (isotropic) pressure correction at the same level of accuracy as the 14-moment approximation method, it provides a better description of the pressure anisotropy. Surprisingly, the Chapman-Enskog iterative procedure leads to equations providing the best overall description in the case of massive, thus non-conformal, fluid dynamics.
\section{Conclusions}
Herein we presented an analysis of the relative importance of the transport coefficients following from various second-order non-conformal viscous hydrodynamics formulations, \textit{i.e.}, Israel-Stewart, 14-moment approximation and Chapman-Enskog method. As a benchmark, we used the new exact solution of the Boltzmann kinetic equation within relaxation-time approximation. We showed explicitly that the inclusion of shear--bulk couplings, which are ignored in Israel-Stewart theory, is crucial for capturing the characteristic evolution of the bulk pressure correction. We also showed that the Chapman-Enskog method leads to a set of equations that provide the best overall description of a non-conformal system. We note, however, that, in conformal case, the best description of the pressure anisotropy can be obtained within anisotropic hydrodynamics framework. We finish with the conclusion that the forms of the transport coefficients entering the relativistic second-order viscous hydrodynamics theory are crucial for the proper description of the quark-gluon plasma within this framework.
\ack
R.R. would like to thank collaborators: G. S. Denicol, W. Florkowski, A. Jaiswal and M. Strickland for useful discussions during preparation of this work. R.R. was supported by Polish National Science Center Grant No.~DEC-2012/07/D/ST2/02125.
\section*{References}

\end{document}